\documentclass{article}
\usepackage{ijcai19}

\usepackage{times}
\usepackage{soul}
\usepackage{url}
\usepackage[hidelinks]{hyperref}
\usepackage[utf8]{inputenc}
\usepackage[small]{caption}
\usepackage{graphicx}
\usepackage{amsmath}
\usepackage{amsthm}
\usepackage{booktabs}
\usepackage{algorithm}
\usepackage{algorithmic}
\usepackage{multirow}
\usepackage{dsfont}
\usepackage[switch]{lineno}
\usepackage{xcolor}

\title{DDIPrompt: Drug-Drug Interaction Event Prediction \\ based on Graph Prompt Learning }

\author{
Yingying Wang$^1$\and
Yun Xiong$^{1}$ \footnote{Corresponding Author}\and
Xixi Wu$^{1}$\and
Xiangguo Sun$^2$\And 
Jiawei Zhang$^3$\\
\affiliations
$^1$Shanghai Key Laboratory of Data Science, School of Computer Science, Fudan University\\
$^2$The Chinese University of Hong Kong,
$^3$University of California, Davis\\
\emails
 22210240301@m.fudan.edu.cn, 
 yunx@fudan.edu.cn,
 21210240043@m.fudan.edu.cn,
xiangguosun@cuhk.edu.hk,
jiawei@ifmlab.org
}

\begin{document}

\maketitle

\begin{abstract}
    Drug combinations can cause adverse drug-drug interactions(DDIs). Identifying specific effects is crucial for developing safer therapies. Previous works on DDI event prediction have typically been limited to using labels of specific events as supervision, which renders them insufficient to address two significant challenges: (1) the bias caused by \textbf{highly imbalanced event distribution} where certain interaction types are vastly under-represented. (2) the \textbf{scarcity of labeled data for rare events}, a pervasive issue where rare yet potentially critical interactions are often overlooked or under-explored due to limited available data. In response, we offer ``DDIPrompt'', an innovative solution inspired by the recent advancements in graph prompt learning. Our framework aims to address these issues by leveraging the intrinsic knowledge from pre-trained models, which can be efficiently deployed with minimal downstream data. Specifically, to solve the first challenge, DDIPrompt features a hierarchical pre-training strategy to foster a generalized and comprehensive understanding of drug properties. It captures intra-molecular structures through augmented links based on structural proximity between drugs, further learns inter-molecular interactions emphasizing edge connections rather than concrete catagories. For the second challenge, we implement a prototype-enhanced prompting mechanism during inference. This mechanism, refined by few-shot examples from each category, effectively harnesses the rich pre-training knowledge to enhance prediction accuracy, particularly for these rare but crucial interactions. Extensive experiments on two benchmark datasets demonstrate DDIPrompt's SOTA performance, especially for those rare DDI events.

\end{abstract}

\section{Introduction}

    Drug safety has always been a concern within the medical community \cite{Effect1,DrugCombine1}. In recent years, there has been an increasing interest in using drug combinations to manage complex diseases, which is promising in improving treatment effectiveness \cite{DrugCombine2}. However, it is crucial to recognize that combining different drugs can occasionally lead to reduced treatment efficacy and even structural modifications of the drugs, posing risks to human health \cite{AdverseDDI1,Effect2}. These negative reactions, known as \underline{d}rug-\underline{d}rug \underline{i}nteraction (DDI) events, carry significant importance when studying drug combinations. Therefore, DDI event prediction becomes a critical research area, aiming to mitigate potential risks and refine treatment outcomes.

DDI event prediction can be modeled as a multi-class edge prediction task on a graph, where drugs are represented as nodes and their interactions as edges \cite{MSAN,Binary1,Binary2,LaGAT}. Graph Neural Networks (GNNs) \cite{GCN,GAT} have emerged as a widely-adopted approach to enhance drug representations for improved predictions. However, existing methods of DDI event prediction, which rely on GNNs, often require a significant amount of supervision information \cite{MRCGNN,GMPNN,HKG-DDIE}, making them dependent on domain expertise for data labeling \cite{R-GCN,Trim-Net}. Besides, these methods struggle to address two inherent challenges in this multi-class edge classification task: (1) the highly imbalanced DDI event distributions. Adverse drug reactions predominantly occur in several common scenarios, leading to model over-fitting of such categories and generating biased predictions. (2) the label scarcity of certain negative drug combinations. The scarcity of some rare DDI events makes it extremely challenging to predict such types, hindering model's effectiveness.

Given the significant biases inherent in the dataset, it is imperative to enrich the knowledge received from pre-training stage, with an expected focus on structural and interactive aspects rather than relying on concrete event categories. Moreover, the label scarcity problem requires an effective approach in the prompting stage that aligns downstream task with pre-training, extracting preserved knowledge for downstream task with minimal supervision. This motivates us to turn to a promising direction, ``pre-train, prompt'', which is originated from the Natural Language Processing (NLP) domain \cite{NLPprompt}. It aims at reformulating downstream task to the pre-training one via a lightweight prompting mechanism while keeping the pre-trained model frozen. This enables the  utilization of preserved knowledge for specific downstream tasks without requiring extensive labels \cite{Prompt1,Prompt2}.

Unfortunately, achieving the above goal is never easy due to two challenges. \textbf{The first one is how to extract unbiased and comprehensive drug knowledge from the highly imbalanced DDI event labels.} Although there have been some works trying to explore the ``pre-train, prompt'' paradigm for GNNs \cite{AllinOne,GraphPrompt,GraphPromptSurvey,GPPT,GPF}, most of them employ pre-training objectives that focus on general graph patterns, like link prediction or node feature reconstruction \cite{VNT}, which is far from sufficient to capture unique knowledge on drug molecules under so imbalanced data. \textbf{Another problem is how to leverage preserved knowledge to facilitate downstream predictions where labeled DDI events are so sparse.} Existing graph prompting works are mostly designed for downstream node or graph classification tasks, presenting a significant semantic gap when compared to DDI event prediction task. Therefore, we need to design an entirely new graph prompting framework to tackle this task.

In light of this, we propose DDIPrompt, a ``pre-train, prompt'' framework with both of the pre-training and prompt learning tailored for few-shot DDI event prediction task. Specifically, to solve the first problem,  we design a hierarchical pre-training method that preserves both intra-molecular and inter-molecular knowledge, ensuring a rich and unbiased understanding of drug properties. During the first pre-training stage, we focus on learning drugs from their distinct structural patterns and also mining potential links between drugs based on structural proximity, fostering a rich structural understanding. In the next pre-training stage, we shift on leveraging the binary interactive relations between drugs instead of concrete categories, mitigating the inherent biases in the label distribution. To solve the second problem, during the prompt tuning stage, DDIPrompt incorporates class prototypes to facilitate the distinct representation of each class, thereby enabling the model to make accurate predictions under few-shot scenarios. These class prototypes, implemented as learnable prompt vectors, are fine-tuned using the available samples, allowing DDIPrompt to adapt to the specific characteristics of each class and improve prediction performance. 
    To summarize, our contributions are as follows:

    \begin{itemize}
        \item Existing DDI event prediction methods are vulnerable to imbalanced distribution and label scarcity of rare events. To overcome these limitations, we propose to apply the ``pre-train, prompt'' paradigm and devise DDIPrompt. To the best of our knowledge, this is the first attempt that applies prompting in the drug domain.

    \item Within the DDIPrompt framework, we devise a novel hierarchical pre-training method that captures both intra-molecular structures and inter-molecular relations, enabling a rich and unbiased understanding of drug properties. 
        
    \item During the prompting stage, a novel prototype-enhanced prompting mechanism is proposed to enable the inference under few-shot learning scenarios.

    \item Extensive experiments on two benchmark datasets show the SOTA performance of DDIPrompt, particularly for those rare events.
    \end{itemize}

\section{Related Work}

\subsection{GNNs for DDI Event Prediction}

GNNs \cite{GCN,GAT,GraphSAGE} have been widely utilized to learn drug features by leveraging two types of graph structural data: 1) intra-molecular: the molecular structure of drugs, and 2) inter-molecular: the relationships between drugs. 

Some methods model each drug as a graph, where atoms are represented as nodes and chemical bonds as edges. GNNs are then applied to learn representations, with optimization performed using DDI event data \cite{SSI-DDI,Trim-Net,GMPNN,Method1}. However, these methods fall short in capturing the rich interactive information between drugs.

To address this limitation, several approaches have been proposed to incorporate the inter-molecular data, thereby enhancing the expressiveness of the models. For instance, model in \cite{SimGNN} applies GNNs on the drug-drug similarity graph. Additionally, some studies first learn representations at the molecular level and then consider DDI event relations to update these representations, resulting in improved binary link prediction \cite{MSAN,dsn-ddi} and edge classification performance \cite{GoGNN,MRCGNN,DANN-DDI}. Other methods introduce additional information from Knowledge Graphs (KGs) to extract knowledge-enriched drug features\cite{KGDDI} and propose multi-scale information fusion techniques for DDI event prediction \cite{MUFFIN,MCFF-MTDDI,MDNN,SumGNN}. Despite their effectiveness, these methods suffer from several weaknesses, including i) the requirement of extensive supervision information for learning GNNs, ii) the inability to handle imbalanced event type distributions, and iii) limited generalization ability for rare events.

\subsection{Graph Prompt Learning}

Prompt-based tuning methods, originated from the NLP domain ~\cite{NLPprompt}, have been widely used to facilitate the adaptation of pre-trained language models to various downstream tasks. Recently, prompt learning has also emerged as a promising tool to cue the downstream tasks in the graph domain \cite{GPPT,GraphPrompt,AllinOne,VNT,prodigy,hgprompt,chen2024prompt}. The pioneering work, GPPT \cite{GPPT} focuses on node classification task and incorporates learnable prompts on graphs. GraphPrompt \cite{GraphPrompt} proposes a uniform prompt design tailored for both node and graph downstream tasks. AllinOne \cite{AllinOne} extends graph prompt learning to encompass prompt tokens, prompt structures, and inserting patterns, introducing a comprehensive prompting framework. Though existing works have explored the ``pre-train, prompt'' paradigm to generalize GNNs, none of them have specifically investigated its application in the context of DDI event prediction. Moreover, it is important to emphasize the limitations inherent in applying existing methods directly to the scenario of DDI event prediction. These methods, typically designed with general pre-training objectives, exhibit a significant gap when tasked with the specialized requirements of downstream DDI prediction, highlighting the need for a customized framework that is specifically tailored to this task.

\begin{figure*}[!t]
    \centering
    \includegraphics[width=18cm]{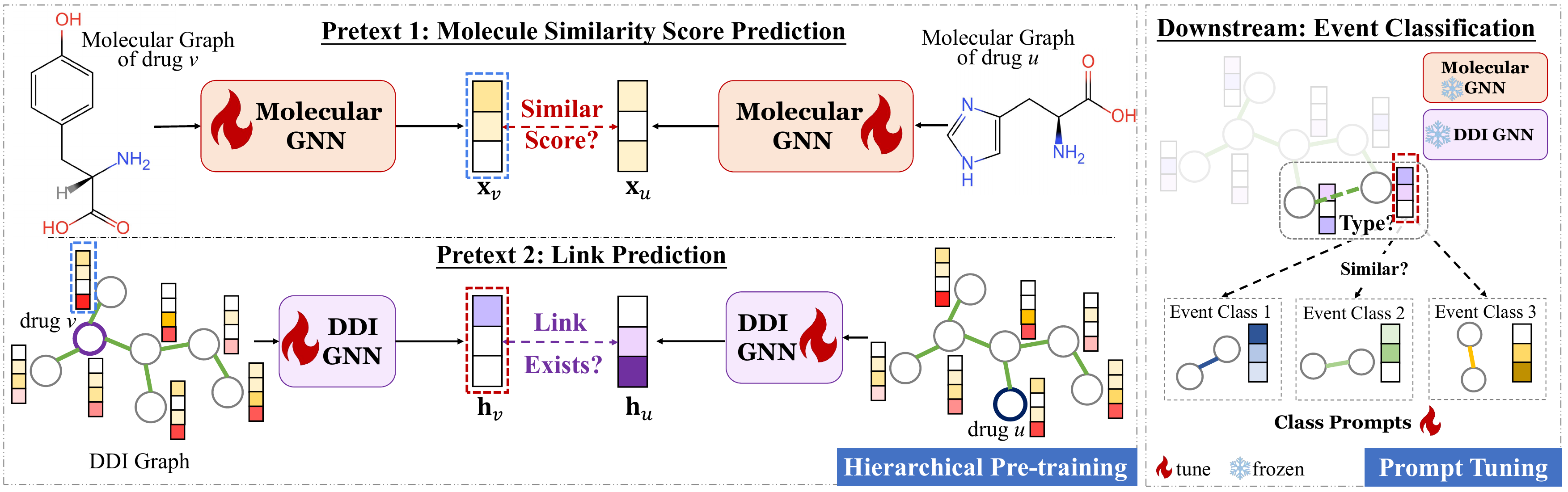}
    \caption{\textbf{Overview of DDIPrompt Framework}. During the pre-training stage, we propose a hierarchical method that first captures structural information of molecules through pairwise similarity prediction and then learns interactive proximity via link prediction. With such a powerful pre-trained model, the prompt learning stage can intelligently finish the downstream prediction task with only a few samples, alleviating previously notorious requirement of labels.}
    \label{fig:model}
\end{figure*}

\section{Methodology}

In this section, we first formulate the DDI event prediction task. Then we elaborate on the pre-training and prompt learning processes within the DDIPrompt framework. The overview of our method is shown in Figure \ref{fig:model}.

\subsection{Task Formulation}
DDI event data can be formulated as a multi-relational DDI event graph $G = (\mathcal{V}, \mathcal{E}, \mathcal{R})$, where $\mathcal{V}=\{ v \}$ denotes the set of drug nodes, $\mathcal{R}=\{ r_k \}_{k=1}^{|\mathcal{R}|}$ denotes the set of relations representing event types, and $\mathcal{E}=\{ (u, v, r_k) | u, v \in \mathcal{V} , r_k \in \mathcal{R}\}$ denotes the set of relational edges among drugs, \textit{i.e.}, DDI events. Additionally, each node $v \in \mathcal{V}$ can be viewed as a drug molecular graph represented by $M_v$, where atoms serve as nodes and bonds serve as edges. Based on $G$ and $\mathcal{M}= \{ M_v \}_{v \in \mathcal{V}}$, our goal is to learn a model which predicts the specific DDI event for each drug pair as $f: \mathcal{V} \times \mathcal{V} \rightarrow \mathcal{R}$.

Before delving into the details of the DDIPrompt framework, we first provide an overview of its two-phase paradigm. In the pre-training stage, we exclusively employ the available structural data, \textit{i.e.}, intra-molecular structures and inter-molecular binary relations, to train GNN models. Subsequently, in the prompt tuning stage, the pre-trained model is frozen, and the task is to predict the event type of the remaining edges given few-shot event samples.

\subsection{Hierarchical Pre-training}
We devise a novel hierarchical pre-training method to address the problem of imbalanced event distribution. Our approach focuses on enhancing existing knowledge through two pretext tasks. We first propose the pretext as predicting molecule similarity score, aiming to augmenting potential links between drugs based on their structural similarities. To further alleviate the inherent bias present in disparate labels, we consider another binary classification task to capture the interactive relations between drugs, \textit{i.e.}, performing link prediction on the DDI graph.



\subsubsection{Pretext 1: Molecule Similarity Score Prediction}
We first represent each drug instance as a molecular graph $M_v$ and employ a GNN parameterized by $\Theta_1$ to capture its distinctive structures as $\mathbf{x}_v = \text{Molecular-GNN}_{\Theta_1}(M_v)$. Specifically, we adopt the dual-view Graph Attention Network proposed in \cite{dsn-ddi}, which consists of a message passing phase and a readout phase.

\begin{itemize}
    \item \textbf{Message Passing Phase} During this phase, each atom $i$ within drug $M_v$ can aggregate messages from neighboring atom nodes $\mathcal{N}_i$ within itself or within its similar drug $M_u$ as follows:

    \begin{equation*}
    \mathbf{x}_i^{(l+1)} = \sigma \left( \sum_{j \in \mathcal{N}_i \cup  \{i\} } \alpha_{ij} \mathbf{W}^{(l)} \mathbf{x}_j^{(l)} + \mathbf{b}^{(l)}  \right), 
    \end{equation*}

    \begin{equation*}
     \alpha_{ij} = \text{Softmax}_j(\mathbf{a}^{(l)T} [ \mathbf{W}^{(l)} \mathbf{x}_i^{(l)} \|   \mathbf{W}^{(l)} \mathbf{x}_j^{(l)}]),
    \end{equation*}

    \noindent where $\mathbf{x}_i^{(l)}$ denotes the representation of atom $i$ in the $l$-th layer, $\mathbf{W}^{(l)}$ is a variable matrix, and $\mathbf{a}^{(l)}, \mathbf{b}^{(l)}$ are trainable vectors in the $l$-th layer. Here, the similar drug $M_u$ is pre-selected based on the Tanimoto Coefficient \cite{TanimotoCoefficient}.

   \item \textbf{Readout Phase} After stacking aforementioned message passing processes for $L$ layers, we obtain atom $i$'s representation $\mathbf{x}_i^{(L)}$. Then, we apply an attentive pooling mechanism to obtain the representation of drug $v$:
   \begin{equation*}
       \mathbf{x}_v = \sigma( \sum_{i \in \mathcal{V}_v} \gamma_i \mathbf{x}_i^{(L)} ), \text{where} \; \gamma_i = \text{Softmax}( \mathbf{c}^T \mathbf{x}_i^{(L)} ).
   \end{equation*}

   Here, $\mathbf{x}_v$ is the learned drug feature, $\mathcal{V}_v$ denotes the node set in molecular graph $M_v$, and $\mathbf{c}$ is a trainable vector.
    
\end{itemize}

Then, we employ a pre-training objective focused on predicting the similarity scores between pairs of drugs to guide the learning of representations $\mathbf{x}_v$ for each drug. The ground-truth score $s_{uv} \in [0,1]$ for a drug pair $(u,v)$ is derived from the Tanimoto coefficient \cite{tanimoto,TanimotoCoefficient}, which is calculated based on laboratory tests. To estimate the similarity score between drugs $u$ and $v$, we introduce a MLP parameterized by $\theta_1$, as $\hat{s}_{uv} = \text{MLP}_{\theta_1}([ \mathbf{x}_u \| \mathbf{x}_v ])$. Then, the learning objective is to minimize the difference between the estimated score $\hat{s}_{uv}$ and the ground-truth score $s_{uv}$ via the Mean Squared Error (MSE) loss function:

\begin{equation*}
    \mathcal{L}_{\text{pre-train}_1} = \frac{1}{|\mathcal{T}_1|} \sum_{(u,v, s_{uv}) \in \mathcal{T}_1}(s_{uv} - \hat{s}_{uv})^2, 
\end{equation*}

\noindent where $\mathcal{T}_1 = \{ (u, v, s_{uv})| u, v \in \mathcal{V} \}$ denotes the training set. This pre-training objective provides a valuable signal for guiding the model to capture meaningful similarities and differences between drugs, thereby enriching the structural knowledge beyond observed drug links.

\subsubsection{Pretext 2: Link Prediction}
Based on observed interactions, we construct a Drug-Drug Interaction (DDI) graph, where the edges are binary-valued to indicate whether there exists an interactive relation.

To encourage mutually enhanced representations between interacted drugs, we introduce an additional GNN encoder implemented by GraphSAGE \cite{GraphSAGE} for feature propagation. Each drug's features are initialized by concatenating its previous intra-molecular representation $\mathbf{x}_v$ with a pre-trained molecular embedding derived from the biomedical knowledge graph \cite{DRKG}, resulting in $\mathbf{h}_v^{(0)} = [ \mathbf{x}_v \| \mathbf{k}_v]$. This initialization enhances the expressiveness of the drug representations. The final representation of drug $v$ is computed as:
\begin{equation*}
    \mathbf{h}_v = \text{DDI-GNN}_{\Theta_2}(\mathbf{h}_v^{(0)},G),
\end{equation*}

\noindent where $\text{DDI-GNN}_{\Theta_2}(\cdot)$ denotes the GNN encoder parameterized by $\Theta_2$, and $G$ represents the DDI graph.

Next, our pre-training objective involves predicting the presence of links between a pair of drugs $(u,v)$. Therefore, we introduce another MLP parameterized by $\theta_2$ to estimate the link probability as $\hat{\mathbf{y}}_{uv} = \text{MLP}_{\theta_2}([ \mathbf{h}_u \| \mathbf{h}_v ])$. The loss is computed via Cross-Entropy (CE) as follows:

\begin{equation*}
    \mathcal{L}_{\text{pre-train}_2} = \frac{1}{|\mathcal{T}_2|} \sum_{(u, v, \mathbf{y}_{uv}) \in \mathcal{T}_2} \text{CE}(\mathbf{y}_{uv}, \hat{\mathbf{y}}_{uv}),
\end{equation*}

\noindent where $\mathcal{T}_2$ denotes the training set for the link prediction task, consisting of both existing edges in the DDI graph as $\{ (u,v, \mathbf{y}_{uv}) | (u,v) \in \mathcal{E}, \mathbf{y}_{uv}=[0, 1]^\top \}$ and randomly sampled negative edges as $\{ (u,v', \mathbf{y}_{uv'}) | (u,v') \notin \mathcal{E}, \mathbf{y}_{uv'}=[1, 0]^\top \}$.

\subsection{Prompt-based Learning and Inference}

After the pre-training stage, we obtain drug $v$'s representation $\mathbf{h}_v$, incorporating both intra-molecular structures and inter-molecular interactive relations. Then, during the prompting stage, given few-shot samples of each event, we aim to infer the specific event types of remaining edges.

To facilitate this inference, we introduce class prompts as prototypes for each event class, which convey class-specific semantics. Instead of tuning the whole pre-trained model, our class prompts are more light-weighted with only a group of vectors containing rich pre-trained knowledge. We utilize the class prompts to reformulate the task as a similarity measure between node representations and the class prompts, aligning with pre-training objectives.


\subsubsection{Prompt Initialization}
Effective initialization of class prompts is crucial for smooth knowledge transfer from pre-trained drug representations to downstream event type classification. For a specific event class $k$, we initialize its class prompt $\mathbf{p}_k$ by computing the mean of embeddings for labeled edges belonging to class $k$:

\begin{equation*}
    \mathbf{p}_k = \frac{1}{|\mathcal{T}^k|} \sum_{(u,v,r_{uv}) \in \mathcal{T}^k} \mathbf{e}_{uv}, \;  \mathbf{e}_{uv} = \text{Mean-Pool}( \{ \mathbf{h}_u , \mathbf{h}_v \}),
\end{equation*}

\noindent where $\mathcal{T}^k = \{ (u, v, r_{uv}) |  r_{uv} = k \}$ denotes the set of labeled edges belonging to type $k$, and $\mathbf{e}_{uv}$ is the edge representation.

\subsubsection{Prompt Tuning}

To learn the distinct features of different classes, we further tune the class prompts by fitting them to the given labels. Specifically, we introduce an MLP to predict the probability of an edge $(u,v)$ belonging to a specific type $k$, as $\hat{r}_{uv}^k = \text{MLP}_{\theta_3}([\mathbf{e}_{uv} \| \mathbf{p}_k ])$, where the edge and class prototype representations are fed to the $\text{MLP}_{\theta_3}(\cdot)$. Subsequently, a similarity measure is applied to transform the resulting vector into a scalar, representing the probability. Optimization is performed via Cross-Entropy loss between the ground-truth labels and predicted types as follows:

\begin{equation*}
    \mathcal{L}_{\text{prompt}} (\{ \mathbf{p}_k \}) = - \sum_{k=1}^K \sum_{(u,v,r_{uv}) \in \mathcal{T}^k}  \mathds{I}(r_{uv}=k) \log \hat{r}_{uv}^k ,
\end{equation*}

\noindent where $\mathds{I}(\cdot)$ denotes an indicator function that returns 1 only if $r_{uv}=k$, and $K$ denotes the total number of classes. Note that this stage is extremely efficient as the number of tunable parameters is $\mathcal{O}(K \times d + L \times d)$, where $d$ refers to the embedding dimension and $L$ refers to the number of layers in $\text{MLP}_{\theta_3}(\cdot)$.

\subsubsection{Prompt-assisted Inference}

After tuning, these class prompts convey class-distinctive semantics and can be directly utilized to perform inference:

\begin{equation*}
    \hat{r}_{uv} = \text{argmax}_{k=1}^K \text{MLP}_{\theta_3}( [\mathbf{e}_{uv} \| \mathbf{p}_k]),
\end{equation*}

\noindent By selecting the type with the highest probability, we determine the predicted event type for the edge $(u,v)$.

\subsection{Complexity Analysis}

Due to space limitations, we move the detailed algorithm process, complexity analysis, and further efficiency study to the Appendix.

\begin{table}[t]
    \small
    \centering
    \caption{\textbf{Event Division of Datasets.} Events are split into common, fewer, and rare three types based on their frequencies.}
    \label{tab:dataset}

    \begin{tabular}{c|ccc}
      \toprule
         & \multicolumn{3}{c}{\textbf{Deng's dataset} \cite{Deng}}  \\
         & Common  & Fewer  & Rare   \\ 
         
       Event Range & \#1 - \#38 & \#39 - \#49 & \#50 - \#65  \\
       Frequencies &  $> 50$ &  $> 15 \; \text{and} \; \leq 50 $ & $<=15$ \\

       \midrule

       & \multicolumn{3}{c}{\textbf{Ryu's dataset} \cite{Ryu}} \\ 
       & Common & Fewer & Rare \\ 
       Event Range & \#1 - \#63 & \#64 - \#75 & \#76 - \#86 \\
       Frequencies &   $> 50$ &  $> 15 \; \text{and} \; \leq 50 $ & $<=15$ \\ 
       
       \bottomrule
    \end{tabular}
\end{table}

\section{Experiment}

\subsection{Experimental Settings}
\textbf{Datasets} We evaluate DDIPrompt on two benchmarks: (1) Deng's dataset \cite{Deng}, which contains a total of 37,159 Drug-Drug Interactions (DDIs) involving 567 drugs and 65 types of DDI events, and (2) DrugBank dataset \cite{Ryu}, which consists of 191,075 DDIs between 1,689 drugs and encompasses 86 types of DDI events. To categorize the DDI events based on their frequencies, we calculate the number of DDI instances associated with each type of event, referred to as \textit{event frequency}. Following the approach outlined in \cite{meta-ddi}, we classify the DDI events into three groups: common events, fewer events, and rare events, as illustrated in Table \ref{tab:dataset}.

\noindent\textbf{Baselines} We compare our DDIPrompt with both representative \textbf{DDI event prediction baselines} (\textit{DeepDDI}, \textit{SSI-DDI}, \textit{MUFFIN}, and \textit{MRCGNN}), and \textbf{few-shot learning methods} (\textit{META-DDI} and \textit{D2PT}). Additionally, we contrast our approach with two strong \textbf{drug pre-training methods}, namely \textit{TransE} and \textit{BioBridge}.
The experimental results are acquired through the integration of our prototype-enhanced prompting mechanism for multi-class inference.
Their brief descriptions are given as follows:
\begin{itemize}
\item \textbf{DeepDDI} \cite{Ryu} pioneers DDI event prediction by designing structural similarity profiles for drugs based on molecular fingerprints.
\item \textbf{SSI-DDI} \cite{SSI-DDI} employs multiple Graph Attention Network (GAT) \cite{GAT} layers on drug molecular graphs to capture molecular structural information for predicting drug pairs.
\item \textbf{MUFFIN} \cite{MUFFIN} integrates embeddings from the Drug Repurposing Knowledge Graph (DRKG) \cite{DRKG} and drug molecular structure features using a bi-level cross strategy. 
\item \textbf{MRCGNN} \cite{MRCGNN} is the state-of-the-art DDI event prediction method. It incorporates contrastive learning for enhanced drug representations.
\item \textbf{META-DDI} \cite{meta-ddi} uses drug chemical structures and interpretable modules for few-shot DDI event prediction.
\item \textbf{D2PT} \cite{D2PT} is a state-of-the-art few-shot learning method on GNNs, capturing long-range dependencies through contrastive learning.
\item \textbf{TransE} \cite{TransE}  is a knowledge graph embedding method that measures the plausibility of triples by transforming it into a distance measurement between head and tail entities.
\item \textbf{BioBridge} \cite{BioBridge} is an advanced multimodal alignment method, which learns from unimodal FMs to establish multimodal behavior based on comprehensive biological knowledge graph.
\end{itemize}

\noindent\textbf{Implementation Details} We implement our DDIPrompt with the PyTorch framework. During the pre-training stage, to implement the molecular graph encoder, for each drug, we select the 10 most similar and 10 most dissimilar drugs based on Tanimoto Score \cite{tanimoto}, forming the training samples for similarity score prediction and graph attention computation. During the prompt tuning stage, we employ the Adam optimizer \cite{adam} with a learning rate of 0.002. Additionally, we initialize the dropout ratio at 0.3 and increase it by 0.05 per training round until it reaches a maximum rate of 0.9. For DDI event prediction, we utilize a 2-layer MLP as the projection head. Details of hyper-parameters setting will be provided in the Appendix.

\noindent \textbf{Evaluation Metrics} We adopt commonly used metrics, including Accuracy, Macro-F1, Macro-Recall, and Macro-Precision. Larger values for all these metrics indicate better performance. To evaluate the effectiveness of the DDIPrompt framework in few-shot scenarios, we employ only \textbf{20\%} of the samples in each event class for prompt tuning (or training data for baseline methods), while reserving the remaining \textbf{80\%} for testing purposes.

\begin{table*}[t]
    \small
   \centering

    \caption{\textbf{Overall Performance Comparison}. The best and second-best results are highlighted with \textbf{bold} and \underline{underline}, respectively.}
    \label{exp:overall}
    \resizebox{\textwidth}{!}{
    \begin{tabular}{c|cc|cccccccc|cc}
       \toprule
       & Dataset & Metric &  DeepDDI & SSI-DDI & MUFFIN & MRCGNN & META-DDIE & D2PT & TransE & BioBridge & DDIPrompt & Improv \% \\
      \midrule
        
     \multirow{16}{*}{\textbf{Deng}} & \multirow{4}{*}{All Events} & Acc & 57.94 & 68.29 & 74.58 & \underline{77.01} & 73.29 & 64.13 & 75.78 & 74.45 & \textbf{78.58} & 2.04 \\
                                  & & F1  & 32.31 & 40.37 & \underline{45.61} & 44.64 & 41.48 & 31.79 & 44.02 & 42.96 & \textbf{48.74} & 6.86 \\
                                  & & Rec & 28.35 & 39.49 & \underline{44.18} & 43.07 & 38.12 & 28.01 & 42.13 & 40.97 & \textbf{45.33} & 2.60 \\
                                 &  & Pre & 33.18 & 45.38 & 47.89 & \underline{48.08} & 45.82 & 35.21 & 46.75 & 45.56 & \textbf{57.65} & 19.90 \\
       \cmidrule{2-13}

     &  \multirow{4}{*}{Common    } & Acc & 59.15 & 67.60 & 75.71 & \underline{78.23} & 75.60 &  66.21 & 76.54 & 75.23 & \textbf{80.41} & 2.79 \\
                                 &  & F1  & 34.26 & 48.06 & 55.80 & \underline{58.24} & 55.24 & 32.86 & 56.29 & 54.94 & \textbf{64.03} & 9.94 \\
                                 &  & Rec & 31.67 & 47.92 & 53.61 & \underline{54.76} & 54.12 & 32.21 & 55.23 & 54.01 & \textbf{61.99} & 13.20 \\
                                 &  & Pre & 37.35 & 49.98 & 60.83 & \underline{65.87} & 59.17 & 34.98 & 67.14 & 65.78 & \textbf{69.17} & 5.01 \\
      \cmidrule{2-13}

      & \multirow{4}{*}{Fewer     } & Acc & 47.18 & 64.40 & 73.40 & \underline{81.89} & 76.85 & 73.00 & 79.21 & 77.83 & \textbf{88.62} & 8.22 \\
                                 &  & F1  & 41.91 & 61.73 & 72.01 & \underline{79.92} & 74.12 & 68.32 & 78.34 & 76.99 & \textbf{87.58} & 9.58 \\
                                 &  & Rec & 41.19 & 60.08 & 70.02 & \underline{77.79} & 72.57 & 67.78 & 77.12 & 75.84 & \textbf{87.35} & 12.29 \\
                                 &  & Pre & 45.44 & 64.21 & 73.81 & \underline{80.36} & 75.43 & 70.21 & 79.56 & 78.12 & \textbf{88.31} & 9.89 \\
       \cmidrule{2-13}

       & \multirow{4}{*}{Rare}   & Acc & 36.36 & 41.17 & 45.97 & 47.27 & \underline{55.13} & 52.16 & 53.18 & 51.83 & \textbf{64.94} & 17.79\\ 

       & & F1 & 31.86 & 38.04 & 44.38 & 43.75 & \underline{51.01} & 48.34 & 49.47 & 48.12 & \textbf{63.60} & 24.69 \\ 

      &  & Rec & 30.62 & 37.21 & 43.21 & 43.73 & \underline{48.28} & 48.17 & 49.23 & 47.87 & \textbf{65.08} & 34.80 \\ 

      &  & Pre & 34.12 & 39.72 & 45.07 & 56.96 & \underline{59.57} & 49.38 & 50.48 & 49.12 & \textbf{72.20} & 21.20 \\
      \midrule

       \multirow{16}{*}{\textbf{DrugBank}} & \multirow{4}{*}{All Events} & Acc & 75.18 & 76.58 & 85.32 & \underline{86.54} & 80.30 & 75.52 & 85.74 & 84.37 & \textbf{89.64} & 2.40 \\
                                 &  & F1  & 54.57 & 51.21 & 65.41 & \underline{73.11} & 61.26 & 53.49 & 72.21 & 71.18 & \textbf{74.89} & 1.05 \\
                                 &  & Rec & 51.78 & 50.02 & 62.25 & \underline{71.04} & 59.30 & 52.17 & 70.23 & 69.26 & \textbf{71.46} & 0.59 \\
                                 &  & Pre & 59.51 & 54.78 & 69.95 & \underline{75.54} & 64.37 & 55.86 & 75.62 & 74.21 & \textbf{82.91} & 8.32 \\
       \cmidrule{2-13}

    &  \multirow{4}{*}{Common    } & Acc & 77.64 & 77.75 & 85.73 & \underline{87.32} & 84.19 & 75.93 & 86.45 & 84.99 & \textbf{89.12} & 0.56 \\
                                  & & F1  & 55.18 & 53.33 & 66.57 & \underline{73.01} & 66.12 & 52.39 & 74.06 & 72.93 & \textbf{79.37} & 8.12 \\
                                 &  & Rec & 53.36 & 52.01 & 63.54 & \underline{72.41} & 63.11 & 51.78 & 71.13 & 69.92 & \textbf{77.18} & 6.29 \\
                                &   & Pre & 60.01 & 55.12 & 74.95 & \underline{74.89} & 67.68 & 55.63 & 77.56 & 76.05 & \textbf{86.54} & 15.25 \\
       \cmidrule{2-13}

     &  \multirow{4}{*}{Fewer     } & Acc & 65.17 & 72.21 & 83.12 & 89.16 & 84.06 & 78.60 & \underline{90.23} & 88.87 & \textbf{96.28} & 6.71 \\
                                 &  & F1  & 62.32 & 71.15 & 78.56 & \underline{88.06} & 79.25 & 74.14 & 86.67 & 85.23 & \textbf{95.75} & 7.51 \\
                                 &  & Rec & 60.61 & 70.03 & 75.93 & \underline{87.94} & 77.43 & 73.01 & 85.12 & 83.73 & \textbf{95.79} & 7.70 \\
                                 &  & Pre & 64.13 & 72.26 & 80.92 & \underline{88.11} & 81.47 & 78.17 & 88.54 & 87.17 & \textbf{96.16} & 7.91 \\
        \cmidrule{2-13}

       & \multirow{4}{*}{Rare} & Acc & 42.43 & 56.17 & 64.28 & 66.67 & \underline{67.58} & 62.38 & 63.12 & 61.89 & \textbf{88.47} & 27.15 \\ 

       & & F1 & 37.23 & 52.37 & 59.78 & 61.21 & \underline{62.21} & 58.23 & 59.47 & 58.12 & \textbf{86.76} & 35.12 \\

      &  & Rec & 34.44 & 50.14 & 57.18 & 58.16 & \underline{59.37} & 55.90 & 57.23 & 55.87 & \textbf{86.61} & 43.47 \\ 

      &  & Pre & 44.09 & 53.19 & \underline{66.36} & 66.17 & 65.58 & 59.32 & 60.48 & 59.12 & \textbf{90.13} & 35.82 \\

       \bottomrule
    \end{tabular}%
    }
\end{table*}

\subsection{Overall Performance Comparison}

\begin{table*}
    \small
    \centering
    \caption{\textbf{Ablation Study}. ``w/o. Mol'', ``w/o. LP'', and ``w/o. Prompt'' refer to the variants that remove the pretext performed on the molecular graph, the pretext performed on the DDI graph, and our prompting mechanism, respectively.}
    \label{exp:ablation}

    \begin{tabular}{cc|cccc|cccc}
     \toprule
       & & \multicolumn{4}{c|}{Deng's Dataset}  & \multicolumn{4}{c}{Ryu's Dataset}  \\
         & & w/o. Mol & w/o. LP & w/o. Prompt & DDIPrompt & w/o. Mol & w/o. LP & w/o. Prompt & DDIPrompt \\ 
      \midrule

      \multirow{4}{*}{All Events} & Acc & 0.7015 & 0.7101 & 0.7201 & \textbf{0.7858} & 0.7841 & 0.7861 & 0.7539 &  \textbf{0.8964} \\ 

      & F1 & 0.3489 & 0.3057 & 0.3742 & \textbf{0.4874} & 0.5267 & 0.4892 & 0.5213 & \textbf{0.7489} \\ 

      & Rec & 0.3215 & 0.2754 & 0.3932 & \textbf{0.4533} & 0.5140 & 0.5879 & 0.4925 & \textbf{0.7146} \\ 

      & Pre & 0.3876 & 0.3556 & 0.3507 & \textbf{0.5765} & 0.6174 & 0.4623 & 0.5932 & \textbf{0.8291} \\ 

      \midrule

      \multirow{4}{*}{Common} & Acc & 0.7119 & 0.7170 & 0.7252 & \textbf{0.8041} & 0.7939 & 0.7978 & 0.7466 & \textbf{0.8912} \\ 

      & F1 & 0.4480 & 0.4539 & 0.5254 & \textbf{0.6403} & 0.6723 & 0.6082 & 0.5038 & \textbf{0.7937} \\ 

      & Rec & 0.4331 & 0.4221 & 0.4872 & \textbf{0.6199} & 0.6517 & 0.5957 & 0.5019 & \textbf{0.7718} \\ 

      & Pre & 0.4718 & 0.5147 & 0.5423 & \textbf{0.6917} & 0.7005 & 0.6848 & 0.5218 & \textbf{0.8654} \\

      \midrule

      \multirow{4}{*}{Fewer} & Acc & 0.7549 & 0.8084 & 0.7773 & \textbf{0.8862} & 0.8817 & 0.9437 & 0.8001 & \textbf{0.9628} \\ 

      & F1 & 0.7562 & 0.8048 & 0.7363 & \textbf{0.8758} & 0.8843 & 0.9397 & 0.7815 & \textbf{0.9575} \\ 

      & Rec & 0.7503 & 0.8054 & 0.7356 & \textbf{0.8735} & 0.8619 & 0.9358 & 0.7908 & \textbf{0.9579} \\

      & Pre & 0.7588 & 0.8206 & 0.7453 & \textbf{0.8831} & 0.8936 & 0.9490 & 0.7947 & \textbf{0.9616} \\ 

      \midrule

      \multirow{4}{*}{Rare} & Acc & 0.4814 & 0.5364 & 0.4297 & \textbf{0.6494} & 0.6858 & 0.7727 & 0.5105 & \textbf{0.8847} \\ 

      & F1 & 0.4461 & 0.5414 & 0.4151 & \textbf{0.6360} & 0.6042 & 0.7438 & 0.5048 & \textbf{0.8676} \\ 

      & Rec & 0.4308 & 0.5435 & 0.4036 & \textbf{0.6508} & 0.5888 & 0.7475 & 0.4998 & \textbf{0.8661} \\ 

      & Pre & 0.4967 & 0.6324 & 0.4499 & \textbf{0.7220} & 0.6227 & 0.8355 & 0.5061 & \textbf{0.9013} \\
      
     \bottomrule
        
    \end{tabular}
    
\end{table*}

The overall performance under weak supervision on both datasets is presented in Table \ref{exp:overall}. Based on these results, we have the following observations:

\begin{itemize}
    \item DDIPrompt consistently outperforms all other baselines on both datasets across all evaluation metrics, demonstrating its superiority. The improvement in rare events is particularly significant, as the F1 score shows a remarkable improvement of 24.69\% compared to the second best method on Deng's dataset. This indicates that the ``pre-train, prompt'' paradigm effectively reduces the reliance on labeled data, enabling DDIPrompt to achieve satisfactory results with only a few samples.
    \item Existing DDI event prediction methods, such as DeepDDI and SSI-DDI, are vulnerable to weak supervision scenarios, as evidenced by their limited effectiveness. These methods heavily rely on a large amount of training samples to achieve optimal performance. However, when confronted with limited data, their performance undergoes a significant decline.
    \item Despite MRCGNN incorporating contrastive learning to address the challenges posed by rare events and the specific design for few-shot scenarios in META-DDIE and D2PT, their performance still falls short of surpassing DDIPrompt. This is because they have not fully captured the rich structural information inherent in drugs to facilitate accurate downstream predictions.
\end{itemize}

\begin{figure*}[!t]
    \centering
    \includegraphics[width=16cm]{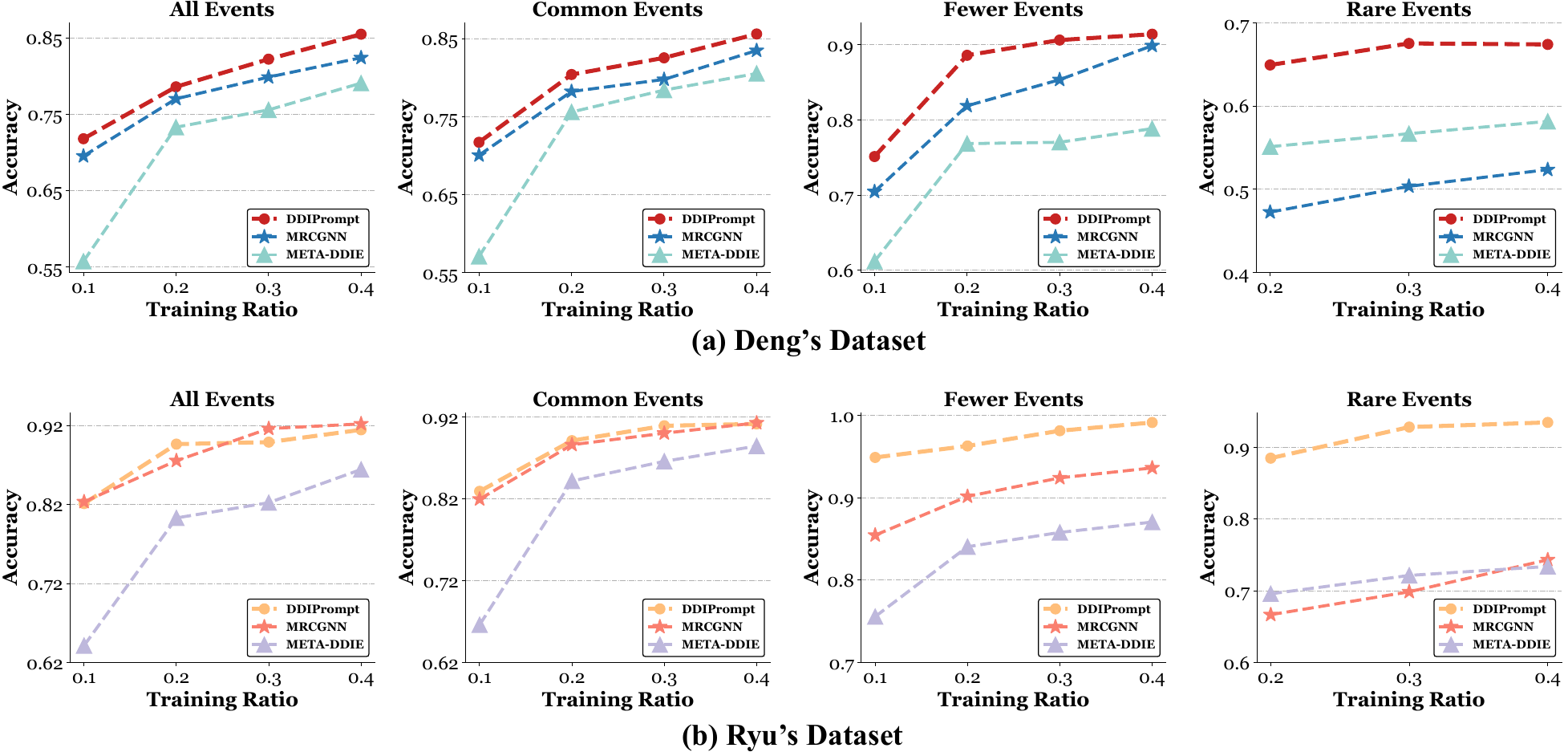}
    \caption{\textbf{Sensitivity Study of Different Methods to Training Sample Size}. }
    \label{fig:sensitivity}
\end{figure*}

\subsection{Ablation Study}
To verify the effectiveness of each component within DDIPrompt, \textit{i.e.}, two pre-training objectives and the prompting mechanism, we consider the following variants: 

\begin{itemize}
\item \textbf{w/o. Mol} To demonstrate the effectiveness of capturing intra-molecular structures, we exclude the pre-training process performed on the molecular graph. Instead, we solely utilize the pre-trained KG embedding as the initial features of drugs in the DDI graph during pre-training.

\item \textbf{ w/o. LP} To illustrate the rationale behind introducing inter-molecular structures, we exclude the pre-training process performed on the DDI graph, \textit{i.e.}, the Link Prediction task.

\item \textbf{w/o. Prompt} To evaluate the effectiveness of our prototype-enhanced prompting process, we replace our prompting mechanism with a MLP-based classifier that fits the provided samples to make predictions.
\end{itemize}

Experimental results are presented in Table \ref{exp:ablation}, where we can observe that removing any part within the DDIPrompt framework consistently leads to performance degradation, highlighting the effectiveness of each component. Both the pre-training objectives and the prototype-based prompting method contribute to the SOTA performance of DDIPrompt.

\subsection{Sensitivity Analysis to Training Sample Size}

In this subsection, we evaluate the sensitivity of our model to the size of training samples used for prompt tuning. While in the overall performance evaluation we uniformly set the ratio to 20\% for prompt tuning, here we vary the ratio from 10\% to 40\% and compare the resulting performance. For comparison, we select the best DDI event prediction baseline, MRCGNN \cite{MRCGNN}, and the best few-shot method META-DDIE \cite{meta-ddi}. The results are shown in Figure \ref{fig:sensitivity}. Note that since some of rare events have frequencies that do not exceed 10, setting the portion to 0.1 would provide no samples for training. Therefore, for rare events, we set the starting point to 0.2. Our findings reveal that DDIPrompt demonstrates strong performance even under extremely weak supervision. Moreover, as the size of the training samples increases, we observe a gradual improvement in performance. These results highlight the robustness and scalability of DDIPrompt, showcasing its ability to achieve satisfactory performance with minimal supervision and further enhance performance as more training data becomes available. In contrast, both MRCGNN and META-DDIE are vulnerable to limited training data and display sensitivity to changes in the size of the training data.

To accommodate space limitations, we relocate the \textbf{Parameters Study} and \textbf{Efficiency Study} to the Appendix.




\section{Conclusion}
In this paper, we propose DDIPrompt, the first "pre-train, prompt" method specifically designed for DDI event prediction. DDIPrompt aims to tackle the challenges posed by distribution imbalance and label deficiency for rare events inherent in DDI data. Our framework incorporates a novel hierarchical pre-training approach that effectively captures the intra-molecular structures and inter-molecular relations, enabling a comprehensive and unbiased understanding of drug properties. Additionally, we propose a prototype-enhanced prompting mechanism that facilitates accurate inference even in few-shot scenarios. Through extensive experiments conducted on two benchmark datasets, we demonstrate that DDIPrompt achieves the SOTA performance.




\clearpage
\newpage

\bibliographystyle{named}

\begin{thebibliography}{}

\bibitem[\protect\citeauthoryear{Asada \bgroup \em et al.\egroup }{2023}]{HKG-DDIE}
Masaki Asada, Makoto Miwa, and Yutaka Sasaki.
\newblock Integrating heterogeneous knowledge graphs into drug--drug interaction extraction from the literature.
\newblock {\em Bioinformatics}, 39(1):btac754, 2023.

\bibitem[\protect\citeauthoryear{Baragaña \bgroup \em et al.\egroup }{2015}]{DrugCombine1}
Beatriz Baragaña, Irene Hallyburton, Marcus C.~S. Lee, et~al.
\newblock A novel multiple-stage antimalarial agent that inhibits protein synthesis.
\newblock {\em Nature}, pages 315--320, 2015.

\bibitem[\protect\citeauthoryear{Bordes \bgroup \em et al.\egroup }{2013}]{TransE}
Antoine Bordes, Nicolas Usunier, Alberto Garcia-Duran, Jason Weston, and Oksana Yakhnenko.
\newblock Translating embeddings for modeling multi-relational data.
\newblock {\em Le Centre pour la Communication Scientifique Directe - HAL - Inria,Le Centre pour la Communication Scientifique Directe - HAL - Inria}, Dec 2013.

\bibitem[\protect\citeauthoryear{Brown \bgroup \em et al.\egroup }{2020}]{NLPprompt}
Tom~B. Brown, Benjamin Mann, and Nick~Ryder et~al.
\newblock Language models are few-shot learners.
\newblock {\em ArXiv}, abs/2005.14165, 2020.

\bibitem[\protect\citeauthoryear{Chen \bgroup \em et al.\egroup }{2021}]{MUFFIN}
Yujie Chen, Tengfei Ma, Xixi Yang, et~al.
\newblock Muffin: multi-scale feature fusion for drug--drug interaction prediction.
\newblock {\em Bioinformatics}, 37(17):2651--2658, 2021.

\bibitem[\protect\citeauthoryear{Chen \bgroup \em et al.\egroup }{2024}]{chen2024prompt}
Xi~Chen, Siwei Zhang, Yun Xiong, Xixi Wu, Jiawei Zhang, Xiangguo Sun, Yao Zhang, Yinglong Zhao, and Yulin Kang.
\newblock Prompt learning on temporal interaction graphs.
\newblock {\em arXiv preprint arXiv:2402.06326}, 2024.

\bibitem[\protect\citeauthoryear{Deac \bgroup \em et al.\egroup }{2019}]{Method1}
Andreea Deac, Yu-Hsiang Huang, Petar Veličković, Pietro Liò, and Jian Tang.
\newblock Drug-drug adverse effect prediction with graph co-attention.
\newblock {\em arXiv: Machine Learning,arXiv: Machine Learning}, May 2019.

\bibitem[\protect\citeauthoryear{Deng \bgroup \em et al.\egroup }{2020}]{Deng}
Yifan Deng, Xinran Xu, Yang Qiu, et~al.
\newblock A multimodal deep learning framework for predicting drug--drug interaction events.
\newblock {\em Bioinformatics}, 36(15):4316--4322, 2020.

\bibitem[\protect\citeauthoryear{Deng \bgroup \em et al.\egroup }{2022}]{meta-ddi}
Yifan Deng, Yang Qiu, Xinran Xu, Shichao Liu, Zhongfei Zhang, Shanfeng Zhu, and Wen Zhang.
\newblock Meta-ddie: predicting drug--drug interaction events with few-shot learning.
\newblock {\em Briefings in Bioinformatics}, 23(1):bbab514, 2022.

\bibitem[\protect\citeauthoryear{Fang \bgroup \em et al.\egroup }{2022}]{GPF}
Taoran Fang, Yunchao Zhang, Yang Yang, Chunping Wang, and Lei Chen.
\newblock Universal prompt tuning for graph neural networks.
\newblock {\em arXiv preprint arXiv:2209.15240}, 2022.

\bibitem[\protect\citeauthoryear{Giacomini \bgroup \em et al.\egroup }{2007}]{Effect1}
Kathleen~M. Giacomini, Ronald~M. Krauss, Dan~M. Roden, et~al.
\newblock When good drugs go bad.
\newblock {\em Nature}, page 975–977, 2007.

\bibitem[\protect\citeauthoryear{Hamilton \bgroup \em et al.\egroup }{2017}]{GraphSAGE}
William~L. Hamilton, Zhitao Ying, and Jure Leskovec.
\newblock Inductive representation learning on large graphs.
\newblock In {\em NIPS}, 2017.

\bibitem[\protect\citeauthoryear{Han \bgroup \em et al.\egroup }{2017}]{Effect3}
Kyuho Han, Edwin~E Jeng, Gaelen~T Hess, et~al.
\newblock Synergistic drug combinations for cancer identified in a crispr screen for pairwise genetic interactions.
\newblock {\em Nature Biotechnology}, page 463–474, 2017.

\bibitem[\protect\citeauthoryear{Han \bgroup \em et al.\egroup }{2023}]{MCFF-MTDDI}
Chen-Di Han, Chun-Chun Wang, Li~Huang, et~al.
\newblock Mcff-mtddi: multi-channel feature fusion for multi-typed drug--drug interaction prediction.
\newblock {\em Briefings in Bioinformatics}, page bbad215, 2023.

\bibitem[\protect\citeauthoryear{Hong \bgroup \em et al.\egroup }{2022}]{LaGAT}
Yue Hong, Pengyu Luo, Shuting Jin, and Xiangrong Liu.
\newblock Lagat: link-aware graph attention network for drug--drug interaction prediction.
\newblock {\em Bioinformatics}, 38(24):5406--5412, 2022.

\bibitem[\protect\citeauthoryear{Huang \bgroup \em et al.\egroup }{2023}]{prodigy}
Qian Huang, Hongyu Ren, Peng Chen, Gregor Kr{\v{z}}manc, Daniel Zeng, Percy Liang, and Jure Leskovec.
\newblock Prodigy: Enabling in-context learning over graphs.
\newblock {\em arXiv preprint arXiv:2305.12600}, 2023.

\bibitem[\protect\citeauthoryear{Ioannidis \bgroup \em et al.\egroup }{2020}]{DRKG}
Vassilis~N. Ioannidis, Xiang Song, Saurav Manchanda, et~al.
\newblock Drkg - drug repurposing knowledge graph for covid-19.
\newblock \url{https://github.com/gnn4dr/DRKG/}, 2020.

\bibitem[\protect\citeauthoryear{Jaaks \bgroup \em et al.\egroup }{2022}]{DrugCombine2}
Patricia Jaaks, Elizabeth~A. Coker, Daniel~J. Vis, et~al.
\newblock Effective drug combinations in breast, colon and pancreatic cancer cells.
\newblock {\em Nature}, pages 166--173, 2022.

\bibitem[\protect\citeauthoryear{Kastrin \bgroup \em et al.\egroup }{2018}]{Binary1}
Andrej Kastrin, Polonca Ferk, and Brane Leskošek.
\newblock Predicting potential drug-drug interactions on topological and semantic similarity features using statistical learning.
\newblock {\em PLOS ONE}, page e0196865, 2018.

\bibitem[\protect\citeauthoryear{Kingma and Ba}{2014}]{adam}
DiederikP. Kingma and Jimmy Ba.
\newblock Adam: A method for stochastic optimization.
\newblock {\em arXiv: Learning,arXiv: Learning}, 2014.

\bibitem[\protect\citeauthoryear{Kipf and Welling}{2017}]{GCN}
Thomas Kipf and Max Welling.
\newblock Semi-supervised classification with graph convolutional networks.
\newblock {\em ArXiv}, abs/1609.02907, 2017.

\bibitem[\protect\citeauthoryear{Lester \bgroup \em et al.\egroup }{2021}]{NLPprompt2}
Brian Lester, Rami Al-Rfou, and Noah Constant.
\newblock The power of scale for parameter-efficient prompt tuning.
\newblock In {\em Conference on Empirical Methods in Natural Language Processing}, 2021.

\bibitem[\protect\citeauthoryear{Li \bgroup \em et al.\egroup }{2021}]{Trim-Net}
Pengyong Li, Yuquan Li, Chang-Yu Hsieh, et~al.
\newblock Trimnet: learning molecular representation from triplet messages for biomedicine.
\newblock {\em Briefings in Bioinformatics}, 22(4):bbaa266, 2021.

\bibitem[\protect\citeauthoryear{Li \bgroup \em et al.\egroup }{2023}]{dsn-ddi}
Zimeng Li, Shichao Zhu, Bin Shao, Xiangxiang Zeng, Tong Wang, and Tie-Yan Liu.
\newblock Dsn-ddi: an accurate and generalized framework for drug--drug interaction prediction by dual-view representation learning.
\newblock {\em Briefings in Bioinformatics}, 24(1):bbac597, 2023.

\bibitem[\protect\citeauthoryear{Lin \bgroup \em et al.\egroup }{2020}]{KGNN}
Xuan Lin, Zhe Quan, Zhi-Jie Wang, Tengfei Ma, and Xiangxiang Zeng.
\newblock Kgnn: Knowledge graph neural network for drug-drug interaction prediction.
\newblock In {\em IJCAI}, volume 380, pages 2739--2745, 2020.

\bibitem[\protect\citeauthoryear{Lin \bgroup \em et al.\egroup }{2022}]{MDF}
Shenggeng Lin, Yanjing Wang, Lingfeng Zhang, et~al.
\newblock Mdf-sa-ddi: predicting drug--drug interaction events based on multi-source drug fusion, multi-source feature fusion and transformer self-attention mechanism.
\newblock {\em Briefings in Bioinformatics}, 23(1):bbab421, 2022.

\bibitem[\protect\citeauthoryear{Liu \bgroup \em et al.\egroup }{2021}]{Prompt1}
Xiao Liu, Yanan Zheng, Zhengxiao Du, et~al.
\newblock Gpt understands, too.
\newblock {\em arXiv: Computation and Language,arXiv: Computation and Language}, 2021.

\bibitem[\protect\citeauthoryear{Liu \bgroup \em et al.\egroup }{2023a}]{DANN-DDI}
Shichao Liu, Yang Zhang, Yuxin Cui, Yang Qiu, Yifan Deng, Zhongfei Zhang, and Wen Zhang.
\newblock Enhancing drug-drug interaction prediction using deep attention neural networks.
\newblock {\em IEEE/ACM Transactions on Computational Biology and Bioinformatics}, 20(2):976--985, 2023.

\bibitem[\protect\citeauthoryear{Liu \bgroup \em et al.\egroup }{2023b}]{D2PT}
Yixin Liu, Kaize Ding, Jianling Wang, et~al.
\newblock Learning strong graph neural networks with weak information.
\newblock In {\em Proceedings of the 29th ACM SIGKDD Conference on Knowledge Discovery and Data Mining}, 2023.

\bibitem[\protect\citeauthoryear{Liu \bgroup \em et al.\egroup }{2023c}]{GraphPrompt}
Zemin Liu, Xingtong Yu, Yuan Fang, and Xinming Zhang.
\newblock Graphprompt: Unifying pre-training and downstream tasks for graph neural networks.
\newblock In {\em Proceedings of the ACM Web Conference 2023}, 2023.

\bibitem[\protect\citeauthoryear{Lyu \bgroup \em et al.\egroup }{2021}]{MDNN}
Tengfei Lyu, Jianliang Gao, Ling Tian, Zhao Li, Peng Zhang, and Ji~Zhang.
\newblock Mdnn: A multimodal deep neural network for predicting drug-drug interaction events.
\newblock In {\em Proceedings of the Thirtieth International Joint Conference on Artificial Intelligence, {IJCAI-21}}, pages 3536--3542, 8 2021.

\bibitem[\protect\citeauthoryear{Ma \bgroup \em et al.\egroup }{2018a}]{SimGNN}
Tengfei Ma, Cao Xiao, Jiayu Zhou, et~al.
\newblock Drug similarity integration through attentive multi-view graph auto-encoders.
\newblock In {\em Proceedings of the Twenty-Seventh International Joint Conference on Artificial Intelligence}, 2018.

\bibitem[\protect\citeauthoryear{Ma \bgroup \em et al.\egroup }{2018b}]{KGDDI}
Tengfei Ma, Cao Xiao, Jiayu Zhou, and Fei Wang.
\newblock Drug similarity integration through attentive multi-view graph auto-encoders.
\newblock In {\em Proceedings of the Twenty-Seventh International Joint Conference on Artificial Intelligence}, Jul 2018.

\bibitem[\protect\citeauthoryear{Maggiora \bgroup \em et al.\egroup }{2014}]{TanimotoCoefficient}
Gerald Maggiora, Martin Vogt, Dagmar Stumpfe, and Jürgen Bajorath.
\newblock Molecular similarity in medicinal chemistry.
\newblock {\em Journal of Medicinal Chemistry}, 57(8):3186--3204, 2014.

\bibitem[\protect\citeauthoryear{Nyamabo \bgroup \em et al.\egroup }{2021}]{SSI-DDI}
Arnold~K Nyamabo, Hui Yu, and Jian-Yu Shi.
\newblock Ssi--ddi: substructure--substructure interactions for drug--drug interaction prediction.
\newblock {\em Briefings in Bioinformatics}, 22(6):bbab133, 2021.

\bibitem[\protect\citeauthoryear{Nyamabo \bgroup \em et al.\egroup }{2022a}]{MGNN1}
Arnold~K Nyamabo, Hui Yu, Zun Liu, et~al.
\newblock Drug–drug interaction prediction with learnable size-adaptive molecular substructures.
\newblock {\em Briefings in Bioinformatics}, Jan 2022.

\bibitem[\protect\citeauthoryear{Nyamabo \bgroup \em et al.\egroup }{2022b}]{GMPNN}
Arnold~K Nyamabo, Hui Yu, Zun Liu, et~al.
\newblock Drug–drug interaction prediction with learnable size-adaptive molecular substructures.
\newblock {\em Briefings in Bioinformatics}, 2022.

\bibitem[\protect\citeauthoryear{Palmer and Sorger}{2017}]{Effect2}
Adam~C. Palmer and Peter~K. Sorger.
\newblock Combination cancer therapy can confer benefit via patient-to-patient variability without drug additivity or synergy.
\newblock {\em Cell}, pages 1678--1691.e13, 2017.

\bibitem[\protect\citeauthoryear{Ryu \bgroup \em et al.\egroup }{2018}]{Ryu}
Jae~Yong Ryu, Hyun~Uk Kim, and Sang~Yup Lee.
\newblock Deep learning improves prediction of drug-drug and drug-food interactions.
\newblock {\em Proceedings of the National Academy of Sciences}, May 2018.

\bibitem[\protect\citeauthoryear{Schick and Schütze}{2021}]{Prompt2}
Timo Schick and Hinrich Schütze.
\newblock Exploiting cloze questions for few shot text classification and natural language inference.
\newblock In {\em Proceedings of the 16th Conference of the European Chapter of the Association for Computational Linguistics: Main Volume}, 2021.

\bibitem[\protect\citeauthoryear{Schlichtkrull \bgroup \em et al.\egroup }{2018}]{R-GCN}
Michael Schlichtkrull, Thomas~N. Kipf, Peter Bloem, Rianne van~den Berg, Ivan Titov, and Max Welling.
\newblock {\em Modeling Relational Data with Graph Convolutional Networks}, page 593–607.
\newblock Jan 2018.

\bibitem[\protect\citeauthoryear{Sun \bgroup \em et al.\egroup }{2022}]{GPPT}
Mingchen Sun, Kaixiong Zhou, Xingbo He, Ying Wang, and Xin Wang.
\newblock Gppt: Graph pre-training and prompt tuning to generalize graph neural networks.
\newblock In {\em Proceedings of the 28th ACM SIGKDD Conference on Knowledge Discovery and Data Mining}, 2022.

\bibitem[\protect\citeauthoryear{Sun \bgroup \em et al.\egroup }{2023a}]{AllinOne}
Xiangguo Sun, Hongtao Cheng, Jia Li, Bo~Liu, and Jihong Guan.
\newblock All in one: Multi-task prompting for graph neural networks.
\newblock In {\em Proceedings of the 29th ACM SIGKDD Conference on Knowledge Discovery and Data Mining}, 2023.

\bibitem[\protect\citeauthoryear{Sun \bgroup \em et al.\egroup }{2023b}]{GraphPromptSurvey}
Xiangguo Sun, Jiawen Zhang, Xixi Wu, Hong Cheng, Yun Xiong, and Jia Li.
\newblock Graph prompt learning: A comprehensive survey and beyond.
\newblock {\em arXiv:2311.16534}, 2023.

\bibitem[\protect\citeauthoryear{Tan \bgroup \em et al.\egroup }{2023}]{VNT}
Zhen Tan, Ruocheng Guo, Kaize Ding, and Huan Liu.
\newblock Virtual node tuning for few-shot node classification.
\newblock In {\em Proceedings of the 29th ACM SIGKDD Conference on Knowledge Discovery and Data Mining}, page 2177–2188, 2023.

\bibitem[\protect\citeauthoryear{Tang \bgroup \em et al.\egroup }{2023}]{DSIL-DDI}
Zhenchao Tang, Guanxing Chen, Hualin Yang, Weihe Zhong, and Calvin Yu-Chian Chen.
\newblock Dsil-ddi: A domain-invariant substructure interaction learning for generalizable drug–drug interaction prediction.
\newblock {\em IEEE Transactions on Neural Networks and Learning Systems}, pages 1--9, 2023.

\bibitem[\protect\citeauthoryear{Tanimoto}{1968}]{tanimoto}
Taffee~T Tanimoto.
\newblock An elementary mathematical theory of classification and prediction, ibm report (november, 1958), cited in: G. salton, automatic information organization and retrieval, 1968.

\bibitem[\protect\citeauthoryear{Tatonetti \bgroup \em et al.\egroup }{2012}]{AdverseDDI1}
Nicholas~P Tatonetti, Guy~Haskin Fernald, and Russ~B Altman.
\newblock A novel signal detection algorithm for identifying hidden drug-drug interactions in adverse event reports.
\newblock {\em Journal of the American Medical Informatics Association}, page 79–85, Jan 2012.

\bibitem[\protect\citeauthoryear{Velickovic \bgroup \em et al.\egroup }{2017}]{GAT}
Petar Velickovic, Guillem Cucurull, Arantxa Casanova, Adriana Romero, Pietro Lio’, and Yoshua Bengio.
\newblock Graph attention networks.
\newblock {\em ArXiv}, abs/1710.10903, 2017.

\bibitem[\protect\citeauthoryear{Vilar \bgroup \em et al.\egroup }{2014}]{Effect4}
Santiago Vilar, Eugenio Uriarte, Lourdes Santana, et~al.
\newblock Similarity-based modeling in large-scale prediction of drug-drug interactions.
\newblock {\em Nature Protocols}, page 2147–2163, 2014.

\bibitem[\protect\citeauthoryear{Wang \bgroup \em et al.\egroup }{2020}]{GoGNN}
Hanchen Wang, Defu Lian, Ying Zhang, et~al.
\newblock Gognn: Graph of graphs neural network for predicting structured entity interactions.
\newblock In {\em Proceedings of the Twenty-Ninth International Joint Conference on Artificial Intelligence}, 2020.

\bibitem[\protect\citeauthoryear{Wang \bgroup \em et al.\egroup }{2021}]{Binary2}
Yingheng Wang, Yaosen Min, Xin Chen, and Ji~Wu.
\newblock Multi-view graph contrastive representation learning for drug-drug interaction prediction.
\newblock In {\em Proceedings of the Web Conference 2021}, WWW '21, page 2921–2933, 2021.

\bibitem[\protect\citeauthoryear{Wang \bgroup \em et al.\egroup }{2024}]{BioBridge}
Zifeng Wang, Zichen Wang, Balasubramaniam Srinivasan, Vassilis~N Ioannidis, Huzefa Rangwala, and Rishita Anubhai.
\newblock Biobridge: Bridging biomedical foundation models via knowledge graphs.
\newblock In {\em International Conference on Learning Representations}, 2024.

\bibitem[\protect\citeauthoryear{Xiong \bgroup \em et al.\egroup }{2023}]{MRCGNN}
Zhankun Xiong, Shichao Liu, Feng Huang, et~al.
\newblock Multi-relational contrastive learning graph neural network for drug-drug interaction event prediction.
\newblock In {\em Proceedings of the AAAI Conference on Artificial Intelligence}, volume~37, pages 5339--5347, 2023.

\bibitem[\protect\citeauthoryear{Yu \bgroup \em et al.\egroup }{2021}]{SumGNN}
Yue Yu, Kexin Huang, Chao Zhang, Lucas~M Glass, Jimeng Sun, and Cao Xiao.
\newblock Sumgnn: Multi-typed drug interaction prediction via efficient knowledge graph summarization.
\newblock {\em Bioinformatics}, 2021.

\bibitem[\protect\citeauthoryear{Yu \bgroup \em et al.\egroup }{2023}]{hgprompt}
Xingtong Yu, Zemin Liu, Yuan Fang, and Xinming Zhang.
\newblock Hgprompt: Bridging homogeneous and heterogeneous graphs for few-shot prompt learning.
\newblock {\em arXiv preprint arXiv:2312.01878}, 2023.

\bibitem[\protect\citeauthoryear{Zhu \bgroup \em et al.\egroup }{2022}]{MSAN}
Xinyu Zhu, Yongliang Shen, and Weiming Lu.
\newblock Molecular substructure-aware network for drug-drug interaction prediction.
\newblock In {\em Proceedings of the 31st ACM International Conference on Information \& Knowledge Management}, pages 4757--4761, 2022.


\end{thebibliography}

\clearpage
\newpage

\appendix
\section{Implementation Details}
\subsection{Hyperparameters Setting}
During the prompt learning stage, we perform grid search to choose the key hyper-parameters. To ensure a comprehensive exploration of the parameter space, groups of hyperparameters are searched together, which is defined as follows:
\begin{itemize}
    \item Learning rate:\{0.001, 0.0015, 0.002, 0.0025\}
    \item Dropout rate:\{0.3, 0.4, 0.5, 0.6\}
    \item Number of epochs:\{100, 200, 300, 400\}
    \item Weight decay:\{1e-4, 3e-4, 5e-4, 7e-4\}
    \item Metric of similarity:\{‘Cosine’,‘Inner’,‘Minkowski’\}
    \item Pooling of edge:\{‘Hadamard’,‘Concat’,‘Sum’,‘Mean’\}
\end{itemize}

Based on the observations from experimental results, the final chosen parameter values are as follows: the learning rate is set to 0.002, the dropout rate is set to 0.3, the number of epochs is set to 300, and the weight decay is set to 3e-4. The similarity metric is selected as the dot product, and the edge pooling method is selected as either Concat or Mean.


\begin{algorithm}[!t]
    \caption{Optimization of Pretext 1 in DDIPrompt}
    \label{alg:pretext1}
    \textbf{Input}: Set of drugs $\mathcal{V}$, set of corresponding drug molecule graphs $\mathcal{M} = \{ M_v \}_{v \in \mathcal{V}}$, and set of drug similarity pairs $\mathcal{T}_1 = \{(u,v,s_{uv}) | u,v \in \mathcal{V} \}$\\
   \textbf{Parameter}:  $\text{Molecular-GNN}_{\Theta_1}(\cdot), \text{MLP}_{\theta_1}(\cdot)$\\
    \textbf{Output}: Drug's molecular embedding matrix $\mathbf{X} = [\mathbf{x}_v]_{v \in \mathcal{V}}$
    \begin{algorithmic}[1] 
        \STATE Initialize $\Theta_1$ in $\text{Molecular-GNN}_{\Theta_1}(\cdot)$, $\theta_1$ in $\text{MLP}_{\theta_1}(\cdot)$
        \WHILE{\textit{not converge}}
        \FOR{ $v \in \mathcal{V}$}
          \STATE Encode drug $v$ as $\mathbf{x}_v = \text{Molecular-GNN}_{\Theta_1}(M_v)$
        \ENDFOR
        \FOR{ $(u,v, s_{uv}) \in  \mathcal{T}_1$}
        \STATE Retrieve $u,v$'s embedding $\mathbf{x}_u, \mathbf{x}_v$ to predict the similarity score as $\hat{s}_{uv} = \text{MLP}_{\theta_1}([\mathbf{x}_u \| \mathbf{x}_v])$
        \STATE Compute $\mathcal{L}_{\text{pre-train}_1}$ based on $s_{uv}$ and $\hat{s}_{uv}$
        \ENDFOR
        \STATE Update $\Theta_1$ and $\theta_1$ based on $\mathcal{L}_{\text{pre-train}_1}$
        \ENDWHILE

    \end{algorithmic}
\end{algorithm}

\begin{algorithm}[!t]
    \caption{Optimization of Pretext 2 in DDIPrompt}
    \label{alg:pretext2}
    \textbf{Input}: DDI interaction graph $G=(\mathcal{V}, \mathcal{E})$, set of link pairs $\mathcal{T}_2 = \{(u,v, \mathbf{y}_{uv}) | u,v \in \mathcal{V} \}$, drug's molecular embedding matrix $\mathbf{X}$, and drug's knowledge embedding matrix $\mathbf{K}$ \\ 
    \textbf{Parameter}: $\text{DDI-GNN}_{\Theta_2}(\cdot)$, $\text{MLP}_{\theta_2}(\cdot)$ \\
    \textbf{Output}: Drug's final embedding matrix $\mathbf{H} = [\mathbf{h}_v]_{v \in \mathcal{V}}$

    \begin{algorithmic}[1]
        \STATE Initialize $\Theta_2$ in $\text{DDI-GNN}_{\Theta_2}(\cdot)$, $\theta_2$ in $\text{MLP}_{\theta_2}(\cdot)$

        \FOR{$v \in \mathcal{V}$}
        \STATE Obtain drug's initial representation $\mathbf{h}_v^{(0)} = [ \mathbf{x}_v \| \mathbf{k}_v]$ 
        \ENDFOR

        \WHILE{\textit{not converge}}

        \FOR{$v \in \mathcal{V}$}
          \STATE Encode drug $v$ as $\mathbf{h}_v = \text{DDI-GNN}_{\Theta_2}(\mathbf{h}_v^{(0)}, G)$
        \ENDFOR

        \FOR{$(u,v, \mathbf{y}_{uv}) \in \mathcal{T}_2$}
        \STATE Retrieve $u,v$'s embedding $\mathbf{h}_u, \mathbf{h}_v$ to predict the link probability as $\hat{\mathbf{y}}_{uv} = \text{MLP}_{\theta_2}( [\mathbf{h}_u \| \mathbf{h}_v])$
        \STATE Compute $\mathcal{L}_{\text{pre-train}_2}$ based on $\mathbf{y}_{uv}$ and $\hat{\mathbf{y}}_{uv}$
        \ENDFOR

        \STATE Update $\Theta_2$ and $\theta_2$ based on $\mathcal{L}_{\text{pre-train}_2}$
        
        \ENDWHILE

    \end{algorithmic}
\end{algorithm}

\subsection{Complexity Analysis}
To provide a comprehensive understanding of the overall efficiency, particularly in the prompt learning stage of the DDIPrompt framework, we present a detailed complexity analysis as follows.

\begin{itemize}
    \item \textbf{Complexity of Pretext 1}: We outline the detailed process in Algorithm \ref{alg:pretext1}. Within each epoch, the graph encoding process in $\text{Molecular-GNN}_{\Theta_1}(\cdot)$ (Line 4) exhibits a complexity of $\mathcal{O}(L_{G_1} |\mathcal{E}_{\text{sum}}|d)$. Here, $L_{G_1}$ represents the number of graph attention layers in $\text{Molecular-GNN}_{\Theta_1}(\cdot)$, $d$ denotes the embedding dimension, and $|\mathcal{E}_{\text{sum}}|$ refers to the sum of edges, \textit{i.e.}, chemical bonds, within each drug molecular graph. Subsequently, the similarity prediction step employs the obtained drug embeddings to predict similarity scores using $\text{MLP}_{\theta_1}(\cdot)$ (Line 7-8). Let $L_{M_1}$ represent the number of layers in $\text{MLP}_{\theta_1}(\cdot)$, the complexity of this step is $\mathcal{O}(L_{M_1}|\mathcal{T}_1|d)$, where $|\mathcal{T}_1|$ corresponds to the number of drug similarity pairs, scaling linearly with the number of drugs $|\mathcal{V}|$. Therefore, the overall complexity within each epoch can be expressed as $\mathcal{O}(|\mathcal{E}_{\text{sum}}| + |\mathcal{V}|)$, scaling \textbf{linearly} with the size of drugs' molecular graphs.


    \item \textbf{Complexity of Pretext 2}: We outline the detailed process in Algorithm \ref{alg:pretext2}. Firstly, graph convolution is performed to obtain the drug's representation (Line 7), resulting in a complexity of $\mathcal{O}(L_{G_2}|\mathcal{E}|d)$, where $L_{G_2}$ denotes the number of convolutional layers, $|\mathcal{E}|$ denotes the number of edges in DDI graph, and $d$ represents the embedding dimension. Subsequently, the prepared link data is utilized for optimization (Line 9-12), contributing to a complexity of $\mathcal{O}(L_{M_2}|\mathcal{E}|d)$, where $L_{M_2}$ signifies the number of layers in $\text{MLP}_{\theta_2}(\cdot)$. Therefore, the overall complexity for performing the link prediction pretext on the DDI graph for one epoch is $\mathcal{O}(L_{G_2}|\mathcal{E}|d+ L_{M_2}|\mathcal{E}|d)=\mathcal{O}(\mathcal{|\mathcal{E}|})$, scaling \textbf{linearly} with the number of edges in DDI graph.
    
    \item \textbf{Complexity of Prompt Learning}: We provide the details of our prompt learning algorithm in Algorithm \ref{alg:prompt}, along with a complexity analysis that encompasses both the prompt learning and inference stages. In the prompt learning stage (Line 4-12), we iterate over each edge within the given set of few-shot samples $\mathcal{T}$ and estimate the probability across $K$ types. This process incurs a complexity of $\mathcal{O}(KL_{M_3}d)$, where $L_{M_3}$ represents the number of layers in $\text{MLP}_{\theta_3}(\cdot)$. Consequently, the overall complexity of the prompt learning stage is $\mathcal{O}(KL_{M_3}|\mathcal{T}|d)$. Since $|\mathcal{T}| \ll |\mathcal{E}|$ , the overall complexity can be considered \textbf{constant}, scaling linearly with the number of provided samples. Next, we analyze the complexity of the inference stage. For each predicted edge in $\mathcal{E} \setminus \mathcal{T}$, we select the maximum probability across $K$ types to obtain predicted label. Therefore, the overall complexity of predicting all edges is $\mathcal{O}(K L_{M_3}|\mathcal{E} \setminus \mathcal{T}|d)$, scales \textbf{linearly} with the number of predicted samples.

\end{itemize}

\begin{algorithm}[!t]
    \caption{Prompt Learning of DDIPrompt}
    \label{alg:prompt}
    \textbf{Input}: DDI graph $G=(\mathcal{V}, \mathcal{E})$, set of few-shot labeled edges $\mathcal{T} = \{ \mathcal{T}^k \}_{k=1}^K$, pre-trained drug embedding matrix $\mathbf{H}$\\ 
    \textbf{Parameter}: Set of prompt vectors $\{ \mathbf{p}_k \}_{k=1}^K$, $\text{MLP}_{\theta_3}(\cdot)$ \\
    \textbf{Output}: Predicted label of edges 

    \begin{algorithmic}[1]
        \FOR{$k = 1,2,..., K$}
          \STATE Initialize $k$-th class prompt $\mathbf{p}_k$  \COMMENT{Prompt Initialization} 
        \ENDFOR

        \WHILE{\textit{not converge}} 
        \FOR{$(u,v, r_{uv}) \in \mathcal{T}$} 

        \FOR{$k=1,2,..., K$}
          \STATE Predict the probability that edge $(u,v)$ belongs to type $k$ as $\hat{r}_{uv}^k$
        \ENDFOR

        \STATE Compute $\mathcal{L}_{\text{prompt}}$ based on $r_{uv}$ and $\{ \hat{r}_{uv}^k \}_{k=1}^K$

        \STATE Update $\{ \mathbf{p}_k \}$ based on $\mathcal{L}_{\text{prompt}}$ \COMMENT{Prompt Tuning}
        \ENDFOR
        \ENDWHILE

        \FOR{$(u,v) \in \mathcal{E} \setminus \mathcal{T}$}
          \STATE $\hat{r}_{uv} = \text{argmax}_{k=1}^K \text{MLP}_{\theta_3}([\mathbf{e}_{uv} \| \mathbf{p}_k]) $ \COMMENT{Inference}
        \ENDFOR
    \end{algorithmic}
\end{algorithm}

Therefore, we can draw conclusions regarding the overall efficiency of the DDIPrompt framework. Firstly, the complexities of both pretexts scale linearly with the size of drug molecular graphs and the DDI graph, respectively. Secondly, the prompt tuning stage exhibits high efficiency. Its complexity can be considered constant, scaling linearly with the number of provided few-shot samples. Finally, the inference stage is also efficient, with a complexity that scales linearly with the number of predicted edges. These characteristics enable the easy implementation, fast computation, and efficient deployment of the DDIPrompt framework.

\section{Supplementary Experiments}

\subsection{Efficiency Study}
The efficiency assessment of DDIPrompt during prompt tuning phase, compared to various baseline models in terms of training time, parameter quantity, and accuracy, is summarized in Table \ref{tab:comparison}. Notably, the parameter quantity for DDIPrompt varies when different edge pooling methods are employed. For instance, the parameter quantity for mean pooling is only half of that for the Concatenation method. The recorded parameter quantity presented here corresponds to the experiments conducted using Concatenation pooling on the Ryu's dataset. On the Deng's dataset, we employed Mean pooling. As shown in Table \ref{tab:comparison}, DDIPrompt demonstrates a competitive balance between these factors. In terms of training time, DDIPrompt achieves remarkable reductions compared to MUFFIN, MRCGNN, and SSI-DDI with optimization percentages of approximately 47.7\%, 35.6\% and 25.2\%, respectively. Similarly, DDIPrompt showcases efficient parameter utilization, surpassing MRCGNN, DeepDDI and META-DDIE with reductions of approximately 30.6\%, 64.7\% and 84.8\%, respectively. This optimization extends to accuracy, with DDIPrompt achieving the SOTA performance compared to baseline models. These results collectively underscore DDIPrompt's significant optimization in training time, parameter quantity, and accuracy, highlighting its superiority across a diverse set of baseline models for DDI event prediction tasks.

\begin{table}[htbp]
  \centering
  \caption{\textbf{Efficiency analysis on Ryu's dataset.} Training time, parameter quantity, and accuracy of DDIPrompt during the prompt tuning phase compared to other baseline models.}
  \label{tab:comparison}
  \resizebox{0.5\textwidth}{!}{%
    \begin{tabular}{lccc}
      \toprule
      \textbf{Method} & \textbf{Training Time (s)} & \textbf{Parameters} & \textbf{Accuracy} \\
      \midrule
      DeepDDI & 15,340 & 3,335,233 & 0.7518 \\
      SSI-DDI & 16,742 & 684,234 & 0.7658 \\
      MUFFIN & 23,819 & 2,059,008 & 0.8532 \\
      MRCGNN & 19,418 & 1,408,915 & 0.8754 \\
      META-DDIE & 101,072 & 7,813,429 & 0.8030 \\
      D2PT & 7,583 & 466,241 & 0.7552 \\
      DDIPrompt & 12,488 & 1,182,038 & 0.8964 \\
      \bottomrule
    \end{tabular}%
  }
\end{table}

\begin{figure}[h]
   \centering  
   \includegraphics[width=0.5\textwidth]{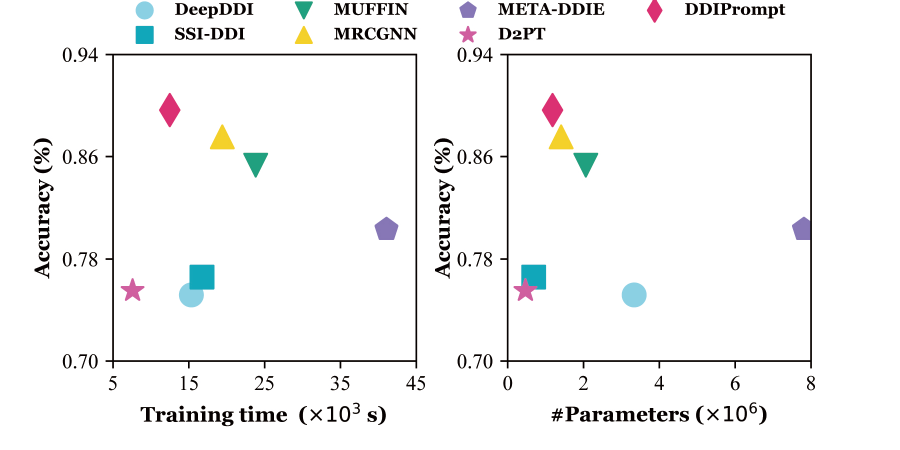}
   \caption{\textbf{Efficiency analysis on Ryu's dataset.}}
   \label{fig:efficiency}
\end{figure}

\subsection{Hyperparameters Study}

We conduct hyperparameter sensitivity analysis on Ryu's dataset to study the influence of two important hyperparameters on the performance of DDIPrompt, \textit{i.e.}, the number of neighboring drugs in $\text{Molecular-GNN}_{\Theta_1}(\cdot)$ and the number of convolutional layers in $\text{DDI-GNN}_{\Theta_2}(\cdot)$

\begin{figure*}[h]
	 \centering  
	 \includegraphics[width=\textwidth]{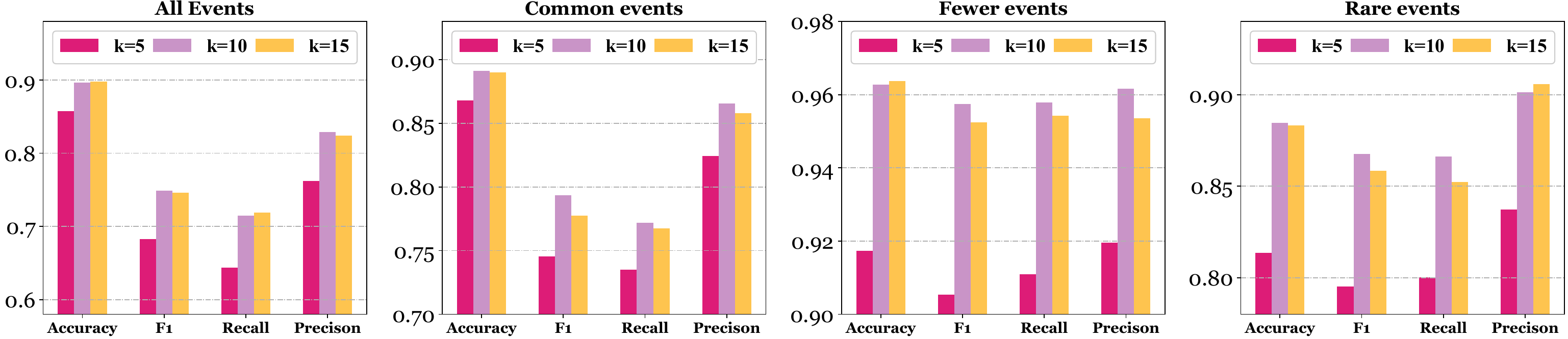}
	 \caption{\textbf{Hyperparameters study on the number of neighboring drugs in $\text{Molecular-GNN}_{\Theta_1}(\cdot)$}. Experimental results on Ryu's dataset are reported.}
      \label{fig:knn}
\end{figure*}

\begin{figure*}[h]
      \centering  
	 \includegraphics[width=\textwidth]{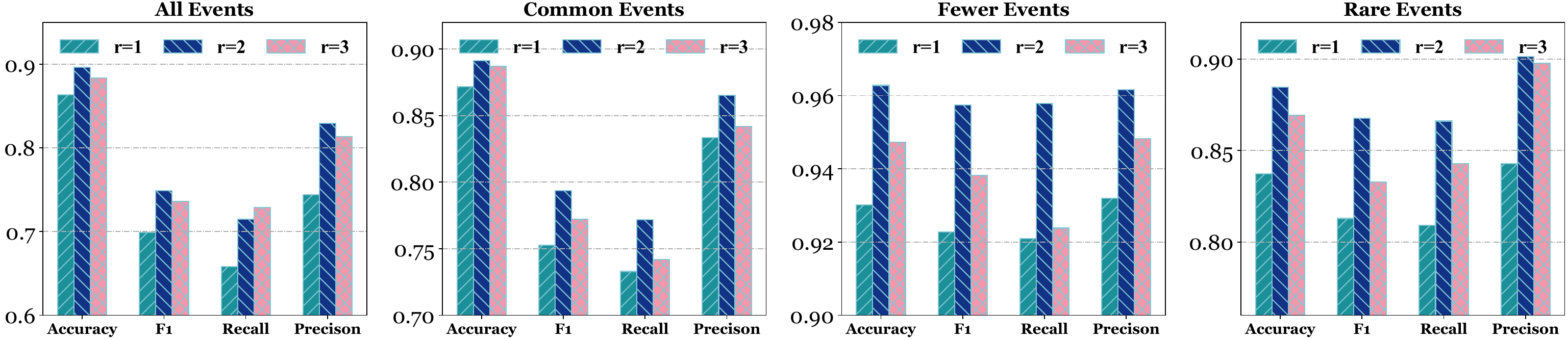}
	 \caption{\textbf{Hyperparameters study on the number of convolutional layers in $\text{DDI-GNN}_{\Theta_2}(\cdot)$}. Experimental results on Ryu's dataset are reported.}
	 \label{fig:layers}
\end{figure*}

\subsubsection{Number of Neighboring Drugs in $\text{Molecular-GNN}_{\Theta_1}(\cdot)$}
Recall that we employ the dual-view Graph Attention Mechanism proposed in \cite{dsn-ddi} to implement $\text{Molecular-GNN}_{\Theta_1}(\cdot)$. This mechanism allows an atom in a drug to receive messages not only from neighboring atoms within the same drug but also from atoms in similar drugs. To construct these similar drug pairs, we set a limit on the number of similar drugs, denoted as $k$, for each drug. This limitation ensures that the number of similar drugs remains reasonable to avoid excessive computational costs. In this investigation, we explore the influence of the $k$ parameter by varying it from 5 to 15. The results on Ryu's dataset are displayed in Figure \ref{fig:knn}. Notably, we observe that the best performance is consistently achieved when $k$ is set to 10 across all scenarios. On the other hand, increasing $k$ to 15 yields only marginal improvements or even performance degradation. Based on these findings, we establish the default setting of $k=10$.

\subsubsection{Number of Convolutional Layers in $\text{DDI-GNN}_{\Theta_2}(\cdot)$}
We also investigate the influence of another parameter, \textit{i.e.}, the number of convolutional layers in $\text{DDI-GNN}_{\Theta_2}(\cdot)$, which is implemented based on GraphSAGE \cite{GraphSAGE}. Specifically, we conduct experiments by varying the number of convolutional layers, denoted as $r$, to be 1, 2, and 3. The results depicted in Figure \ref{fig:layers} demonstrate that setting $r$ to 2 already achieves optimal performance. Further increasing the value of $r$ may lead to the over-smoothing issue, without yielding improved performance. Therefore, we establish the number of convolutional layers in $\text{DDI-GNN}_{\Theta_2}(\cdot)$ as 2.

\end{document}